\documentclass[aps,prd,reprint,nofootinbib,longbibliography]{revtex4-1}

\usepackage{amsmath}
\usepackage{mathtools}
\usepackage{graphicx}
\usepackage[normalem]{ulem}
\usepackage[com pat=1.1.0]{tikz-feynman}
\usepackage{tikz}
\usepackage{aas_macros}
\usetikzlibrary{external}
\usepackage{comment}
\usetikzlibrary{arrows}
\usepackage[colorlinks=true, allcolors=blue]{hyperref}

\newcommand\barparena[1]{\overset{%
   \scriptscriptstyle(-)}{#1}}

\begin{document}

\title{Time-dependent and quasi-steady features of fast neutrino-flavor conversion}

\author{Hiroki Nagakura}
\email{hiroki.nagakura@nao.ac.jp}
\affiliation{Division of Science, National Astronomical Observatory of Japan, 2-21-1 Osawa, Mitaka, Tokyo 181-8588, Japan}
\author{Masamichi Zaizen}
\affiliation{Faculty of Science and Engineering, Waseda University, Tokyo 169-8555, Japan}

\begin{abstract}
Despite the theoretical indication that fast neutrino-flavor conversion (FFC) ubiquitously occurs in core-collapse supernova and binary neutron star merger, the lack of global simulations has been the greatest obstacle to study their astrophysical consequences. In this {\it Letter}, we present large-scale ($50 {\rm km}$) simulations of FFC in spherical symmetry by using a novel approach. We effectively rescale the oscillation scale of FFC by reducing the number of injected neutrinos in the simulation box, and then extrapolate back to the case of the target density of neutrinos with a convergence study. We find that FFC in all models achieves quasi-steady state in the non-linear regime, and its saturation property of FFC is universal. We also find that temporal- and spatial variations of FFC are smeared out at large radii due to phase cancellation through neutrino self-interactions. Finally, we provide a new diagnostic quantity, ELN-XLN angular crossing, to assess the non-linear saturation of FFC. 
\end{abstract}
\maketitle

{\em Introduction.}---Core-collapse supernova (CCSN) and binary neutron star merger (BNSM) are unique laboratories to prove neutrino and nuclear physics. The promise of future multi-messenger observations exhibits possibilities of placing constraints on properties of nuclear matter, neutrino oscillation parameters, and also revealing the origin of heavy elements in our universe.

Neutrino quantum kinetics has received attention recently, but there still remain many open questions \cite{2010ARNPS..60..569D,2016NuPhB.908..366C,2020arXiv201101948T}. The fast neutrino-flavor conversion (FFC), which is induced by refractive effects of neutrino-neutrino self-interactions, is of the greatest interest \cite{2005PhRvD..72d5003S}. FFC can occur in deep inner core of CCSN and BNSM, and may shuffle neutrino flavors almost instantly. Sensitive dependence of weak interactions on neutrino flavors suggests that FFC has an impact on fluid dynamics and nucleosynthesis \cite{2020PhRvD.102j3015G,2021PhRvL.126y1101L,2022arXiv220316559J}. Recent theoretical indications that FFC occurs in CCSN \cite{2021PhRvD.103f3033A,2021PhRvD.104h3025N} and BNSM \cite{2017PhRvD..95j3007W} environments further boost the motivation of its detailed investigation.

Study of FFC requires solving quantum kinetic equation (QKE). Numerical simulations are useful tools to explore the non-linear dynamics. However, the tremendous disparity of time and length scales makes global simulations a formidable challenge. In fact, only local-box FFC simulations have been carried out so far \cite{2017JCAP...02..019D,2019PhRvD.100b3016M,2020PhRvD.102j3017J,2020PhLB..80035088M,2020PhRvD.102f3018B,2021PhRvL.126f1302B,2021PhRvD.104j3003W,2021PhRvD.103f3001M,2021PhRvD.104j3023R,2021PhRvD.104h3035Z,2021PhRvD.104l3026D,2022JCAP...03..051A,2022PhRvD.105d3005S,2022arXiv220505129B,2022arXiv220506282R}. On the other hand, FFC is very sensitive to angular distributions of neutrinos, which can be determined only with global geometrical effects. This exhibits that local simulations are not capable of providing astrophysical consequences of FFC, which has been an enduring obstacle. In this {\it Letter}, we propose a novel approach to address this issue; we rescale a self-interaction potential by reducing neutrino number density systematically, and then the results are extrapolated to the no-rescaling case with a convergence study. We demonstrate that this approach has a capability to provide new insights of non-linear FFC dynamics on $>10 {\rm km}$ scales.

\begin{figure*}
    \includegraphics[width=\linewidth]{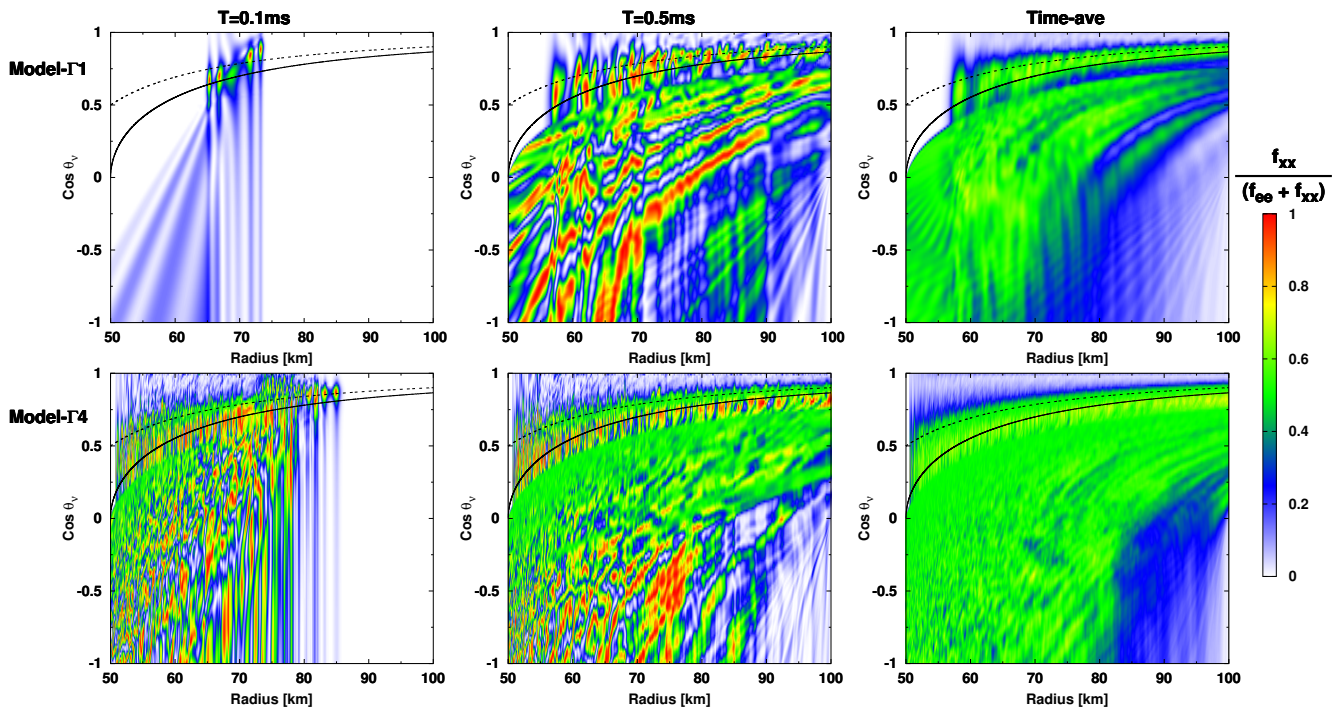}
    \caption{All plots show $f_{xx}/(f_{ee}+f_{xx})$ as functions of radius and $\cos \theta_{\nu}$. Top and bottom panels show results of Model-$\Gamma$1 and Model-$\Gamma$4, respectively. The left and middle panels display the result at $t=0.1 {\rm ms}$ and $0.5 {\rm ms}$, respectively. The right panels depict time-averaged distributions in a quasi-steady state phase ($0.3 {\rm ms} \le t \le 0.5 {\rm ms}$). The black solid- and dashed lines represent trajectories of neutrinos emitted in the direction of $\cos \theta_{\nu} = 0$ (perpendicular to the radial direction) and $\cos \theta_{\nu} = 0.5$ (ELN crossing point), respectively, at the inner boundary ($50 {\rm km}$).
}
    \label{graph_Radi_vs_angular}
\end{figure*}

{\em Method and model.}---
In this study, we use a neutrino transport code GRQKNT; the capability of our code is described in another paper \cite{2022arXiv220604098N}. We work in flat spacetime, spherical symmetry, and neglecting matter-interactions on both collision term and neutrino Hamiltonian. The resultant QKE can be written as,
\begin{equation}
  \begin{split}
& \frac{\partial \barparena{f}}{\partial t}
+ \frac{1}{r^2} \frac{\partial}{\partial r} ( r^2 \cos \theta_{\nu}  \barparena{f} )  - \frac{1}{r \sin \theta_{\nu}} \frac{\partial}{\partial \theta_{\nu}} ( \sin^2 \theta_{\nu} \barparena{f}) \\
&  = - i [\barparena{H},\barparena{f}].
  \end{split}
\label{eq:BasEq}
\end{equation}
In the expression, we use $c = 1$ where $c$ denotes the light speed. $\barparena{f}$ denotes the density matrix of neutrinos (anti-neutrinos). $t, r,$ and $\theta_{\nu}$ represent time, radius, and neutrino flight direction, respectively. We note that the determinant of the spatial metric is not unity in spherical polar coordinate (the second term in left side of Eq.~\ref{eq:BasEq}), and there is an advection term of neutrino angular direction in momentum space (the third term in left side of the equation). These are corrections of QKE from plane-parallel to spherically-symmetric geometry (see \cite{2022arXiv220604098N} for more details). $\barparena{H}$ corresponds to the oscillation Hamiltonian with vacuum and self-interaction contributions.

We assume that electron-type neutrinos and their anti-partners (hereafter, we refer them as $\nu_e$ and $\bar{\nu}_e$, respectively) are emitted outwards (inward) at the sphere of inner (outer) boundary with different angular distributions in momentum space. Their angular distributions are determined with the following equation,
\begin{equation}
\barparena{f}_{ee} = \langle \barparena{f}_{ee}\rangle \biggl( 1 + \barparena{\beta}_{ee} ( \cos \theta_{\nu} - 0.5 ) \biggr) \hspace{4mm} \cos \theta_{\nu } \ge 0,
\label{eq:iniang_in}
\end{equation}
at inner boundary, and
\begin{equation}
\barparena{f}_{ee} = \langle \barparena{f}_{ee} \rangle \times \eta \hspace{4mm} \cos \theta_{\nu } < 0,
\label{eq:iniang_out}
\end{equation}
at outer boundary. $\langle \barparena{f}_{ee} \rangle$ and $\barparena{\beta}_{ee}$ are control parameters, which are directly associated with the number density and anisotropy of angular distributions of neutrinos, respectively. In this study, number density of $\nu_e$ ($n_{\nu}$) is set to be $6 \times 10^{32} {\rm cm^{-3}}$ at the inner boundary ($50 {\rm km}$), which corresponds to $L_{\nu} \sim 4 \times 10^{52} {\rm erg/s}$ for $E_{\rm ave} \sim 12 {\rm MeV}$, where $L_{\nu}$ and $E_{\rm ave}$ denotes the $\nu_e$ luminosity and average energy, respectively. We assume $\langle f_{ee} \rangle = \langle \bar{f}_{ee} \rangle$, $\beta_{ee} = 0$, and $\bar{\beta}_{ee} = 1$.
The parameter $\eta$ in Eq.~\ref{eq:iniang_out} represents the diluteness of incoming neutrinos emitted from outer boundary, which is set to be $\eta=10^{-6}$.

\begin{figure*}
    \includegraphics[width=\linewidth]{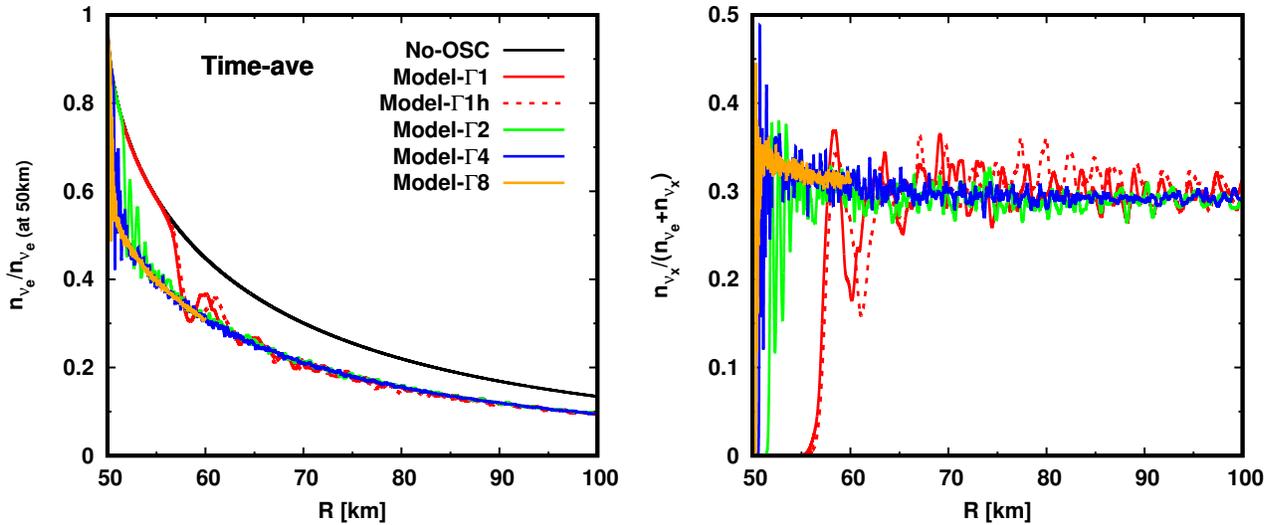}
    \caption{Radial profiles of time-averaged neutrino number density in a quasi-steady state. In the left panel, we show $n_{\nu_e}$ normalized by that at $50 {\rm km}$. For comparison, the result without FFC is also shown as a black solid line. In the right panel, we display $n_{\nu_x}/(n_{\nu_e} + n_{\nu_x})$, which corresponds to a useful metric to see the degree of flavor conversion.
}
    \label{graph_RadproNumdens}
\end{figure*}

In the setup, the oscillation wavelength of FFC at $50 {\rm km}$ is subcentimeter. The required radial resolution is, hence, $\sim 0.1 {\rm cm}$, illustrating that unfeasible computational resources are needed for global simulations. We tackle this issue in the following way. First, we introduce a new parameter, $\Gamma$, which represents a reduction factor of $n_{\nu}$. It effectively rescales the oscillation scale ($\sim 10^4$ times larger than the target one as shown below), which makes $> 10 {\rm km}$ simulations tractable.
Second, we run multiple simulations with different choice of $\Gamma$; in this {\it Letter} we study four cases: $\Gamma=10^{-4}$ (Model-$\Gamma$1), $2 \times 10^{-4}$ (Model-$\Gamma$2), $4 \times 10^{-4}$ (Model-$\Gamma$4), and $8 \times 10^{-4}$ (Model-$\Gamma$8). To see the impact of angular resolution, we also run another simulation (Model-$\Gamma$1h), in which $\Gamma$ is set to be the same as Model-$\Gamma$1 but the angular resolution is twice higher.

We cover a spatial domain of $50 {\rm km} \le r \le 100 {\rm km}$ except for Model-$\Gamma$8. Although Model-$\Gamma$8 covers the narrow spatial domain ($50 {\rm km} \le r \le 60 {\rm km}$), it corresponds to the highest $n_{\nu}$ among our models, and therefore the model is worthy to extrapolate our results to the case with $\Gamma=1$. We deploy 128 angular grid points in our simulations, and only Model-$\Gamma$1h has 256 angular points. In the radial direction, we employ uniform grids with 24576 (for Model-$\Gamma$1 and Model-$\Gamma$1h), 49152 (for Model-$\Gamma$2), 98304 (for Model-$\Gamma$4), and 49152 (for Model-$\Gamma$8) points. It should be stressed that these large number of grids are necessary to resolve FFC (an oscillation wavelength is resolved by $\gtrsim 10$ radial grid points).

We impose a Dirichlet boundary condition for outgoing neutrinos ($\cos{\theta_{\nu}}>0$) at the inner boundary, and for incoming neutrinos ($\cos{\theta_{\nu}}<0$) at the outer one. In the opposite directions, we impose a free-streaming boundary condition. To prepare the initial condition, we run the simulations without FFC until the system settles into a steady state. In FFC simulations, we follow the time evolution up to $0.5 {\rm ms}$ ($0.12 {\rm ms}$ only for Model-$\Gamma$8), which is long enough to establish a quasi-steady state. We work in two-flavor approximations, and employ vacuum potential with $\Delta m^2 = 2.5 \times 10^{-6} {\rm eV^2}$, $\theta_{\rm mix} = 10^{-6}$ and $E_{\nu}=12 {\rm MeV}$, where $\Delta m^2$ and $\theta_{\rm mix}$ denote squared mass difference of neutrinos, mixing angle, and neutrino energy, respectively. We note that the vacuum oscillation is only important to trigger FFC, and it does not affect non-linear regimes of FFC in our models. We confirm by linear analysis that $\Gamma=10^{-4}$ is large enough so that the fast mode dominates over the slow one.

\begin{figure*}
    \includegraphics[width=\linewidth]{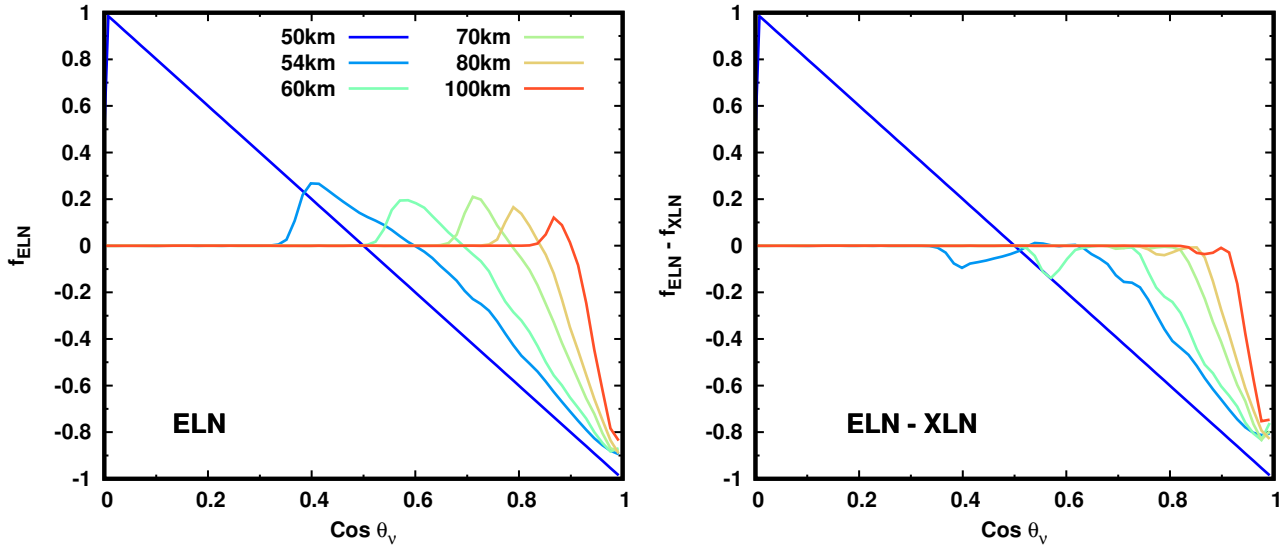}
    \caption{Time-averaged angular distributions of ELN (left) and ELN - XLN (right) at different radii for Model-$\Gamma$4. Time average is taken in $0.3 {\rm ms} \le t \le 0.5 {\rm ms}$. The vertical scale is normalized so that $f_{ELN}$ in the direction of $\cos {\theta_{\nu}}=0$ at $r=50 {\rm km}$ becomes unity.
}
    \label{graph_angdistriFixedR}
\end{figure*}

{\em Results.}---
Figure~\ref{graph_Radi_vs_angular} shows color maps of $f_{xx}$ normalized by $f_{ee}+f_{xx}$ as functions of radius and neutrino angle. The black solid line in each panel portrays the radial trajectory of neutrinos emitted perpendicular to radial direction ($\cos {\theta_{\nu}}=0$) at the inner boundary. This exhibits a transition to forward-peaked angular distributions of neutrinos.

As shown in Fig.~\ref{graph_Radi_vs_angular}, FFC commonly occurs in our models (appearance of $\nu_x$ is a sign of flavor conversion). In the vicinity of inner boundary, however, no strong flavor conversions occur (see, e.g., $50 {\rm km} < r \lesssim 65 {\rm km}$ in the top left panel of Fig.~\ref{graph_Radi_vs_angular}), whereas the region becomes narrower with increasing $n_{\nu}$ (see the bottom panels). This is attributed to the fact that the growth of FFC becomes more rapid with increasing $n_{\nu}$\footnote{In the case without reduction of $n_{\nu}$, the width of corresponding region is only $\sim 20 {\rm cm}$; see the left panel of Fig.~11 in \cite{2022arXiv220604098N}.}.

Once neutrinos, initially emitted in the radial direction from the inner boundary, arrive at a certain radius, flavor structures in all neutrino angles are disorganized (see, e.g., bottom left panel of Fig.~\ref{graph_Radi_vs_angular}), despite the fact that neutrinos traveling in non-radial directions have not reached yet (since the propagation speed of neutrinos with respect to radial direction is proportional to $\cos {\theta_{\nu}}$). This indicates that the flavor conversion in non-radial directions is not a consequence of spatial advection from the inner region (where FFC has already been well developed), but rather local angular-couplings of FFC. This also exhibits that neutrinos emitted from the outer boundary can experience strong flavor conversion. Since the incoming neutrinos are very dilute, their contribution to neutrino self-interaction potential is very minor, suggesting that the flavor conversion is passively induced by outgoing ones. These incoming neutrinos, possessing finite flavor off-diagonal components of the density matrix, advect inward, which facilitates the growth of FFC in the linear regime.

Strong flavor conversion occurs even in the case of low $n_{\nu}$ models at late times (see the top middle panel of Fig.~\ref{graph_Radi_vs_angular}), and we find that the system eventually achieves a quasi-steady state. One of the striking results in this study is that the degree of flavor conversion does not hinge on $n_{\nu}$ in the quasi-steady phase. This trend is more visible in time-averaged distributions. We compute the time-averaged $f$ by integrating over the time of $0.3 {\rm ms} \le t \le 0.5 {\rm ms}$; the results are shown in right panels of Fig.~\ref{graph_Radi_vs_angular}. Fig.~\ref{graph_RadproNumdens} also displays the radial profiles of time-averaged number density of $\nu_e$ and the ratio of $n_{\nu_x}$ to $n_{\nu_e}+n_{\nu_x}$ in the left and right panel, respectively, for all models (for Model-$\Gamma$8, we compute the time-averaged $f$ in the time range of $0.06 {\rm ms} \le t \le 0.12 {\rm ms}$). Both figures illustrate that the degree of flavor mixing is universal. It should also be mentioned that the angular resolutions in our simulations does not compromise the time-averaged profile (see the red dashed-line in Fig.~\ref{graph_RadproNumdens} displaying the result of Model-$\Gamma$1h).
The result of Model-$\Gamma$8, that corresponds to the model with the highest spatial resolution and the modest $\Gamma$, also strengthens our conclusion. As shown in Fig.~\ref{graph_RadproNumdens}, the results of other models clearly approach to Model-$\Gamma$8 with increasing $n_{\nu}$. This lends confidence to our claim that the case of $\Gamma=1$ (no reduction of $n_{\nu}$) can be studied from these results.

Temporal- and spatial variations of FFC are vigorous even after the system reaches a quasi-steady state, indicating that the system never achieves the exact steady state. On the other hand, these fluctuations become mild with increasing radius (see the region of $80 {\rm km} \lesssim r \lesssim 100 {\rm km}$ in Fig.~\ref{graph_Radi_vs_angular}). In the sense of classical neutrino transport, this feature is at odds, because temporal variations generated at the inner region can be sustained in the free-streaming region. The suppression of inhomogeneity is, hence, dictated by quantum effects. Since neutrinos propagating along different trajectories have random temporal variations, these variations can be canceled each other through self-interactions. We note that this is different from the so-called "kinematic decoherence" (see, e.g., \cite{2010ARNPS..60..569D}), albeit similar mechanism. In fact, the vacuum oscillation is nothing to do with them, and more importantly, the flavor equipartition has been almost achieved at the inner region. Our result suggests that temporal variations of FFC occurring deep inner core would be smeared out during the flight in the free streaming region.

Finally, we analyze angular distributions of neutrinos in quasi-steady state, which provides a new insight to understand non-linear saturation of FFC. In this analysis, we pay a special attention to angular distributions of ELN and XLN (heavy-neutrino lepton number). We find that the time-averaged ELN angular distribution subtracted by that of XLN (hereafter, we refer to it as ELN-XLN angular distribution) is a key quantity. As shown in the right panel of Fig.~\ref{graph_angdistriFixedR}, ELN-XLN angular crossings, which exists at the inner boundary, disappear at large radii, whereas ELN crossings still remains (see left panel of Fig.~\ref{graph_angdistriFixedR}). Our result suggests that FFC evolves towards eliminating ELN-XLN angular crossings in the linear phase, and then the growth of FFC is saturated when the ELN-XLN crossing vanishes. Although we postpone the detailed study of the mechanism \cite{TMP_Zaizen_Nagakura}, our interpretation is supported by a linear analysis. As is well known, ELN crossing is a good indicator of occurrence of FFC. However, this is true only if heavy leptonic neutrinos and anti-partners are zero or they have the same angular distributions each other. In non-linear regime of FFC, heavy leptonic neutrinos and their antipartners substantially emerge with different angular distributions. Hence, their contribution needs to be taken into account. Since the traceless part of density matrix is proportional to $(f_{ee} - \bar{f}_{ee})-(f_{xx} - \bar{f}_{xx})$ in two flavor system (see, e.g., \cite{2017PhRvL.118b1101I}), ELN-XLN crossings are more natural quantities to characterize the stability of FFC than ELN ones.

{\em Conclusions.}---
This paper presents large-scale ($50 {\rm km}$) simulations of FFC in spherical symmetry by using GRQKNT code \cite{2022arXiv220604098N}. We make simulations tractable by reducing neutrino number density ($n_{\nu}$) at the inner sphere. By running multiple simulations with changing the reduction rate of $n_{\nu}$, we can study temporal- and quasi-steady features of FFC on $> 10 {\rm km}$ spatial scale. One striking result found in this study is that the time-averaged distributions of FFC is less sensitive to $n_{\nu}$. We also find that the fluctuations are smoothed out at the outer radii as a consequence of phase cancellation through neutrino self-interactions. These features would not be changed in the case without reduction of $n_{\nu}$. Our simulations suggest that ELN-XLN angular distribution is a key quantity to characterize non-linear saturation of FFC; in fact, time-averaged ELN-XLN angular crossings disappear in the region where the non-linear saturation is achieved.

For comprehensive understanding of FFC dynamics, much work lies ahead. It is intriguing to study the dependence of initial angular distribution of neutrinos in more realistic situation, for instance, under the three neutrino flavor treatment. In fact, the three flavor treatment is mandatory to assess occurrences of FFC \cite{2021PhRvD.103f3013C} and to study influences of $\mu$ and $\tau$ neutrinos on non-linear regime of FFC precisely \cite{2021PhRvD.104j3023R,2021PhRvD.104b3011S,2021PhRvD.104h3035Z,2022arXiv220506272C}. We also plan to incorporate effects of neutrino emission, absorption, and momentum-exchanged scatterings, which characterize FFC dynamics at optically thick and the neutrino decoupling region. In CCSN and BNSM environments, these semi-transparent regions are not spherically symmetric and gravity is strong, suggesting that general relativistic multi-dimensional neutrino transport is indispensable. It is also an intriguing question how we can combine our results with those obtained by some existing long-term simulations of FFC \cite{2020PhRvD.102f3018B,2021PhRvD.104j3003W,2021PhRvD.104j3023R}. As discussed in this {\it Letter}, effects of phase cancellation due to different neutrino trajectories would be a key element to improve these local models. Although we leave these broader tasks to future work, the present study exhibits possibilities of performing large-scale FFC simulations in currently operating supercomputer facilities. This is an important progress towards studying observational consequences of FFC in CCSN and BNSM.

\section{Acknowledgments}
The authors thank Lucas Johns, Chinami Kato, Sherwood Richers, George Fuller, Taiki Morinaga, and Shoichi Yamada for useful comments and discussions. The numerical simulations are carried out by using "Fugaku" and the high-performance computing resources of "Flow" at Nagoya University ICTS through the HPCI System Research Project (Project ID: 210050, 210051, 210164, 220173, 220047). and XC50 of CfCA at the National Astronomical Observatory of Japan (NAOJ). For providing high performance computing resources, Computing Research Center, KEK, and JLDG on SINET of NII are acknowledged. This work is supported by the Particle, Nuclear and Astro Physics Simulation Program (Nos. 2022-003) of Institute of Particle and Nuclear Studies, High Energy Accelerator Research Organization (KEK). MZ is supported by a JSPS Grant-in-Aid for JSPS Fellows (No. 22J00440) from the Ministry of Education, Culture, Sports, Science, and Technology (MEXT) in Japan.
\bibliography{bibfile}

\end{document}